\documentclass[twocolumn,pra,aps,superscriptaddress,showpacs,10pt]{revtex4-2}
\usepackage{amsmath}
\usepackage{amssymb}
\usepackage{color}
\usepackage{graphicx}
\usepackage{indentfirst}
\usepackage{qcircuit}
\usepackage{subfig}
\usepackage[left=2.5cm, right=2.5cm, top=2.5cm, bottom=2.5cm]{geometry}
\usepackage[ruled,linesnumbered]{algorithm2e}
\usepackage[hidelinks]{hyperref}
\graphicspath{{fig/}}

\newcommand{\ket}[1]{\left|#1\right\rangle}

\newcommand{\argmax}{\operatornamewithlimits{argmax}}

\begin{document}

\title{Unbiased Quantum Phase Estimation}

\author{Xi Lu}
\affiliation{School of Mathematical Science, Zhejiang University, Hangzhou, 310027, China}
\affiliation{State Key Lab. of CAD\&CG, Zhejiang University, Hangzhou, 310058, China}

\author{Hongwei Lin}
\email{hwlin@zju.edu.cn}
\affiliation{School of Mathematical Science, Zhejiang University, Hangzhou, 310027, China}
\affiliation{State Key Lab. of CAD\&CG, Zhejiang University, Hangzhou, 310058, China}

\begin{abstract}
    Quantum phase estimation algorithm (PEA) is one of the most important algorithms in early studies of quantum computation.
    It is also a key for many other quantum algorithms, such as the quantum counting algorithm and the Shor's integer factorization algorithm.
    However, we find that the PEA is not an unbiased estimation, which prevents the estimation error from achieving an arbitrarily small level.
    In this paper, we propose an unbiased phase estimation algorithm (UPEA) based on the original PEA, and study its application in quantum counting.
    We also show that a maximum likelihood post-processing step can further improve its robustness.
    In the end, we apply UPEA to quantum counting, and use an additional correction step to make the quantum counting algorithm unbiased.
\end{abstract}
\maketitle

\section{Introduction}

Early quantum algorithms are mostly based on two algorithms, the Grover's search algorithm~\cite{Grover1997} and the quantum Fourier transformation (QFT)~\cite{Kitaev1995, Shor1997}.
The quantum phase estimation algorithm (PEA)~\cite{Kitaev1995} is one of the most important applications of QFT, as well as a key for many other quantum algorithms, such as the quantum counting algorithm~\cite{QuantumCounting1998} and the Shor's integer factorization algorithm~\cite{Shor1997}.
The PEA based order finding sub-procedure is considered as the source of the exponential speedup of the Shor's algorithm.

Though PEA was proposed over 20 years ago, it is still a research hotspot in recent years~\cite{Svore2013,Wiebe2016,Wie2019,Suzuki2019,Grinko2021}.
Besides, the \textit{iterative phase estimation algorithm} (IPEA)~\cite{Dobsicek2008,dobvsivcek2007arbitrary,mosca1998hidden} is a more NISQ(noise-intermediate scale quantum)-friendly variant for PEA.
With a certain stretagy of selecting $\phi$, IPEA can be identity to PEA~\cite{Dobsicek2008}.

\begin{figure}[ht]
    \centering
    $$\Qcircuit @C=0.7em @R=0.4em {
        \lstick{\ket{0}} & \qw & \gate{H} & \qw & \qw & \qw & \ctrl{4} & \multigate{3}{\mathrm{QFT}^\dagger} & \meter & \cw \\
        \lstick{\vdots} & & & & & \vdots & & \nghost{QFT^\dagger} & \vdots & & \\
        \lstick{\ket{0}} & \qw & \gate{H} & \qw & \ctrl{2} & \qw & \qw & \ghost{\mathrm{QFT}^\dagger} & \meter & \cw \\
        \lstick{\ket{0}} & \qw & \gate{H} & \ctrl{1} & \qw & \qw & \qw & \ghost{\mathrm{QFT}^\dagger} & \meter & \cw \\
        \lstick{\ket{0}} & \qw{/^n} & \gate{H^{\otimes n}} & \gate{U}& \gate{U^2} & \qw & \gate{U^{2^{t-1}}} & \qw & \qw & \qw
    }$$
    \caption{ The quantum circuit of PEA. }
    \label{fig:circuit-PEA}
\end{figure}
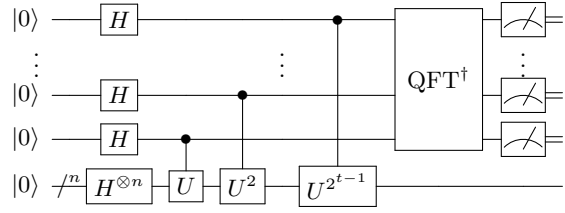

Given a quantum circuit that performs unitary transformation $U$, and an eigenstate $\ket{\psi}$ of $U$ such that 
\begin{equation}
    U\ket{\psi}=e^{2\pi i\varphi}\ket{\psi},
\end{equation}
the \textit{phase estimation algorithm} (PEA)~\cite{Kitaev1995} provides an efficient way to estimate $\varphi$.
The QFT-based form of PEA~\cite{Kitaev1995,Shor1997} uses the circuit shown in \autoref{fig:circuit-PEA}.

Let the integer-encoded measurement result be $s$, then
\begin{equation}
    \tilde{\varphi} = \frac{s}{T}
\end{equation}
is an estimation of $\varphi$, where $T=2^t$, and $t$ is the number of qubits involved in QFT in \autoref{fig:circuit-PEA}.
The result obeys the following distribution~\cite{qcqi},
\begin{gather}
    P_{PEA}(\tilde{\varphi}|\varphi) = \left(\frac{\sin(T\pi(\tilde{\varphi}-\varphi))}{T\sin(\pi(\tilde{\varphi}-\varphi))}\right)^2, \nonumber
    \\ \tilde{\varphi}\in\left\{0,\frac{1}{T},\frac{2}{T},\cdots,\frac{T-1}{T}\right\}
    \label{eq:dist-PEA}
\end{gather}

\begin{figure}
    \centering
    \includegraphics[width=.45\textwidth]{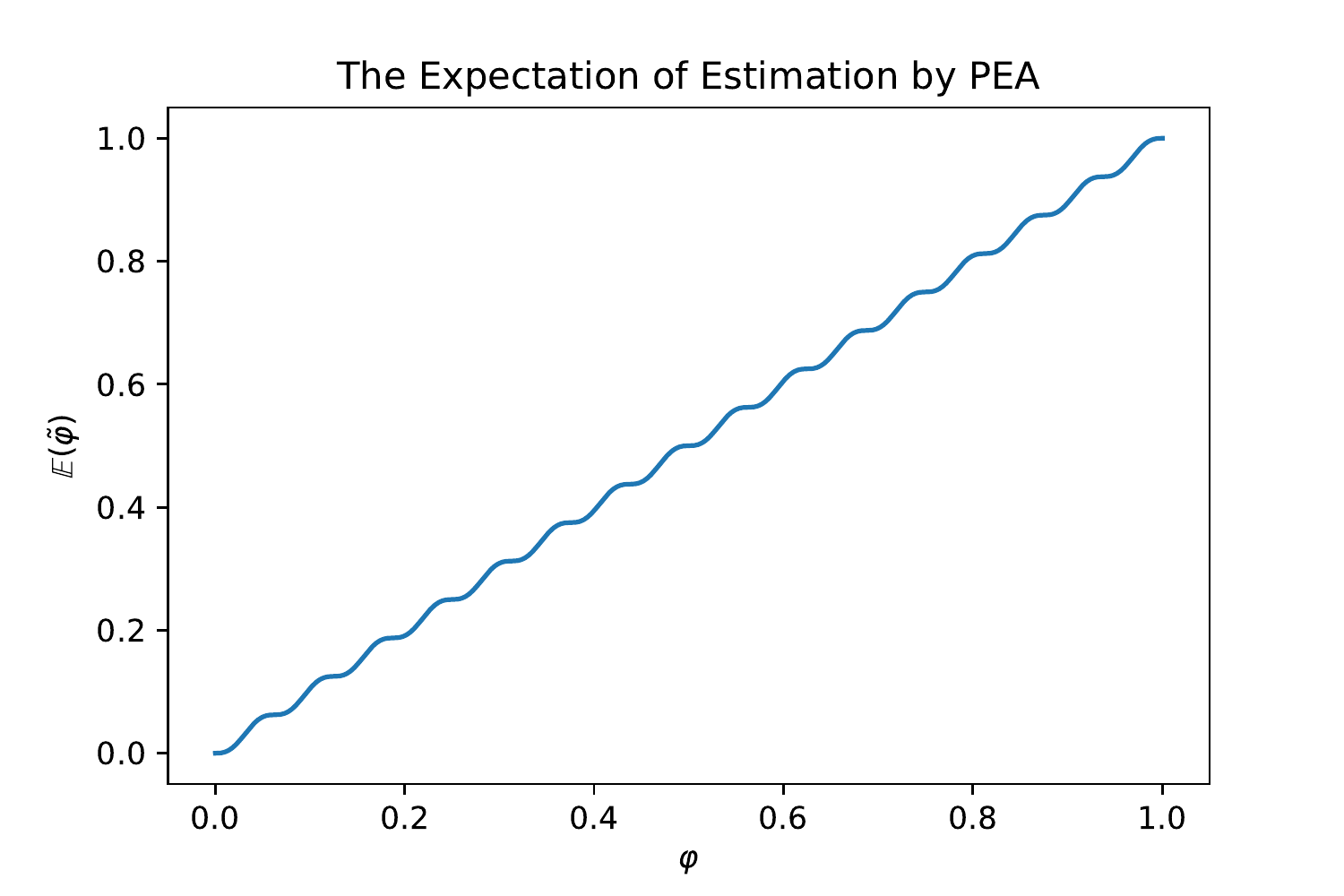}
    \caption{The theoretical expectation of $\tilde{\varphi}$ deducted from Eq.~\eqref{eq:dist-PEA}, in which $T=16$. An unbiased estimation should satisfy $\mathbb{E}(\tilde{\varphi})=\varphi$.}
    \label{fig:expectation-pea}
\end{figure}

In \autoref{eq:dist-PEA}, the estimation is accurate when $\varphi$ is an integer multiplication of $T^{-1}$, and shows the biggest noise when $\varphi$ is a half integer multiplication of $T^{-1}$.
Using the theoretical distribution, it is not hard to find that PEA is biased periodically, as shown in \autoref{fig:expectation-pea}.
The bias can prevent the estimation error from reaching an arbitrarily small level by repetitions.

In this paper, we propose an unbiased phase estimation algorithm (UPEA) based on the original PEA, and study its application in quantum counting.
We also show that a maximum likelihood post-processing step can further improve its robustness.
In the end, we apply UPEA to quantum counting, and use an additional correction step to make the quantum counting algorithm unbiased.

The meanings of UPEA are evident.
The accuracy of PEA decreases linearly about $T^{-1}$.
But on real quantum computers, $T$ is limited by the qubit numbers, decoherence time, gate fidelity and so on.
And the accuracy of simple repetitions of PEA is limited by the bias.
In comparison, repetitions of UPEA can bring down the accuracy to arbitrary level due to the unbiasedness.

\section{Unbiased Phase Estimation}

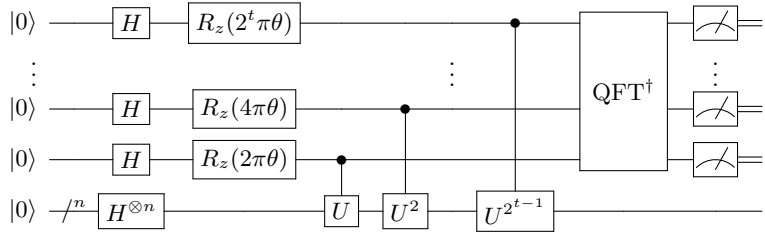
\begin{figure*}
    \centering
    $$\Qcircuit @C=1.0em @R=0.5em {
        \lstick{\ket{0}} & \qw & \gate{H} & \gate{R_z(2^t\pi\theta)} & \qw & \qw & \qw & \ctrl{4} & \multigate{3}{\mathrm{QFT}^\dagger} & \meter & \cw \\
        \lstick{\vdots} & & & & & & \vdots & & \nghost{QFT^\dagger} & \vdots & & \\
        \lstick{\ket{0}} & \qw & \gate{H} & \gate{R_z(4\pi\theta)} & \qw & \ctrl{2} & \qw & \qw & \ghost{\mathrm{QFT}^\dagger} & \meter & \cw \\
        \lstick{\ket{0}} & \qw & \gate{H} & \gate{R_z(2\pi\theta)} & \ctrl{1} & \qw & \qw & \qw & \ghost{\mathrm{QFT}^\dagger} & \meter & \cw \\
        \lstick{\ket{0}} & \qw{/^n} & \gate{H^{\otimes n}} & \qw & \gate{U}& \gate{U^2} & \qw & \gate{U^{2^{t-1}}} & \qw & \qw & \qw
    }$$
    \caption{ The quantum circuit of UPEA. }
    \label{fig:ciruit-UPEA}
\end{figure*}

The key idea of UPEA is intuitive.
In each individual run of PEA, we uniformly randomly choose $\theta\in[0,1]$, or $\theta\sim U(0,1)$.
The circuit of UPEA is shown in \autoref{fig:ciruit-UPEA}.
The quantum state right before the $\mathrm{QFT}^\dagger$ gate is,
\begin{equation}
    \sum_{j=0}^{T-1} e^{2\pi ij(\varphi+\theta)}\ket{j}\ket{\psi}.
\end{equation}

Let $s$ be the measurement result, then $s/T$ is an estimation of the quantity $\varphi+\theta$ in UPEA.
Note that $\theta$ is a known classical parameter, our estimation of $\varphi$ is given by,
\begin{equation}
    \tilde{\varphi} = \frac{s}{T} - \theta.
\end{equation}

Different from PEA, the output $\tilde{\varphi}$ in UPEA is a continuous random variable.
Here we calculate the probability density function $\rho_{UPEA}(\tilde{\varphi})$.
A necessary condition for obtaining an estimation $\tilde{\varphi}$ is,
\begin{equation}
    \theta \in \frac{1}{T}\mathbb{Z} - \varphi,
\end{equation}
which appears exactly $T$ times in the interval $[0,1)$.
Thus,
\begin{align}
    \rho_{UPEA}(\tilde{\varphi};\varphi) 
    = & \sum_{\theta \in \frac{1}{T}\mathbb{Z} - \varphi} P_{PEA}(\tilde{\varphi}+\theta|\varphi+\theta) 
    \\ = & \frac{\sin^2(T\pi(\tilde{\varphi}-\varphi))}{T\sin^2(\pi(\tilde{\varphi}-\varphi))}.
\end{align}

We use bias and mean absolute error (MAE) to quantify the performance of PEA and UPEA.
Using the theoretical distribution in \autoref{eq:dist-PEA}, the bias and MAE of PEA is given by,
\begin{align}
    B_{PEA}(\varphi) &= \sum_{\tilde{\varphi}} d(\tilde{\varphi},\varphi) P_{PEA}(\tilde{\varphi}|\varphi),\\
    M_{PEA}(\varphi) &= \sum_{\tilde{\varphi}} \left|d(\tilde{\varphi},\varphi)\right| P_{PEA}(\tilde{\varphi}|\varphi),
\end{align}
where the signed circular distance $d$ is defined as,
\begin{equation}
    d(\tilde{\varphi},\varphi) := \left[\operatornamewithlimits{argmin}_{\varphi'\in\tilde{\varphi}+\mathbb{Z}}|\varphi'-\varphi|\right] - \varphi.
\end{equation}

The bias and MAE of UPEA is given by,
\begin{align}
    B_{UPEA}(\varphi)
    = & \int_0^1 B_{PEA}(\varphi+\theta)\operatorname{d}\theta
    \\ = & \int_{-1/2}^{1/2} B_{PEA}(\theta)\operatorname{d}\theta,
    \\
    M_{UPEA}(\varphi)
    = & \int_0^1 M_{PEA}(\varphi+\theta)\operatorname{d}\theta
    \\ = & \int_{-1/2}^{1/2} M_{PEA}(\theta)\operatorname{d}\theta,
\end{align}
where all functions about $\varphi$ and $\tilde{\varphi}$ are periodically extended from $[0,1]$ to $\mathbb{R}$.

Observing that $P(\tilde{\varphi}|\varphi)=P(-\tilde{\varphi}|-\varphi)$ and $d(\tilde{\varphi},\varphi)=-d(-\tilde{\varphi},-\varphi)$, $B_{PEA}(\theta)$ is an odd function about $\theta$, thus $B_{UPEA}(\varphi)=0$.
This proves the unbiasedness of our UPEA algorithm.
Also, $M_{UPEA}(\varphi)$ is constant over $\varphi$.

\begin{figure*}
    \centering
    \subfloat[The bias calculated by simulating PEA for $2^{16}$ times.]{
        \includegraphics[width=.45\textwidth]{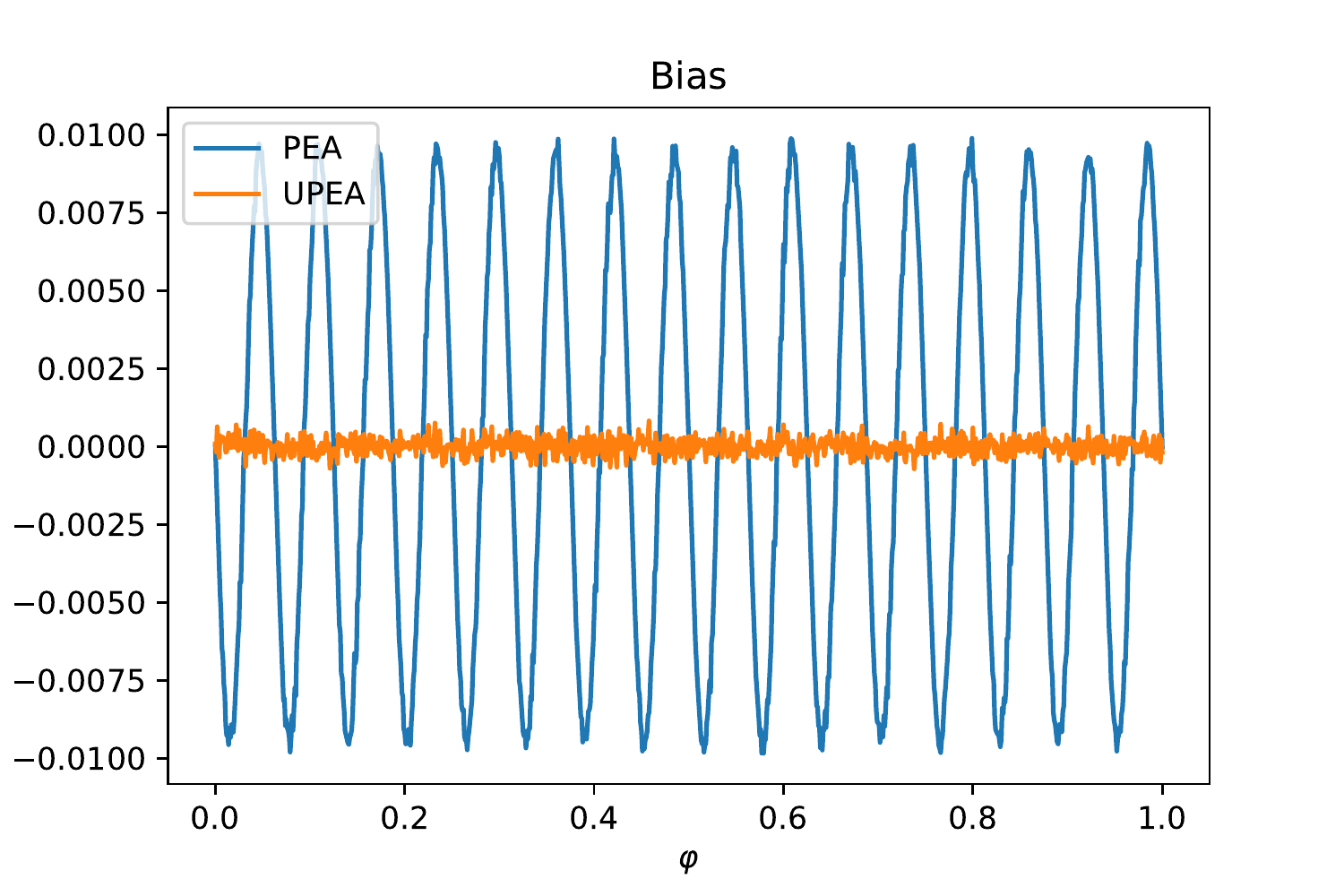}
    }\quad
    \subfloat[The MAE calculated by simulating PEA for $2^{16}$ times.]{
        \includegraphics[width=.45\textwidth]{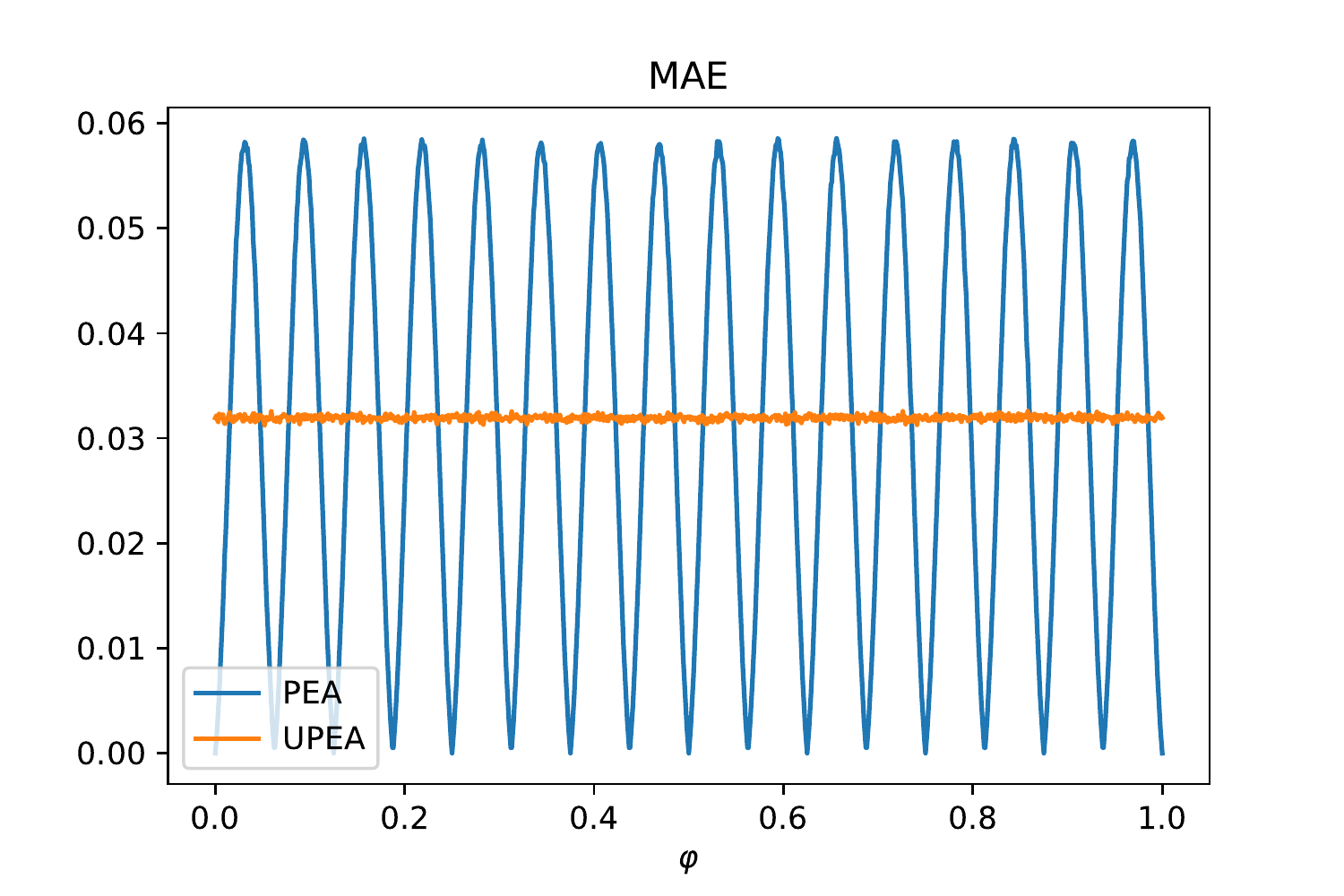}
    }
    \caption{The bias and MAE of PEA and UPEA by sampling the theoretical distribution for $2^{16}$ times, with parameter $T=16$. The $x$-axis stands for $\varphi$, and the $y$-axis stands for bias or MAE.}
    \label{fig:bias_mae_upea}
\end{figure*}

We simulate the UPEA for $2^{16}$ times for analyzing the bias and MAE, and the results are illustrated in \autoref{fig:bias_mae_upea}.
Consistent with our theoretical analysis, the bias is close to zero everywhere, and the MAE is nearly constant.

In the end, it should be mentioned that since the function $P_{PEA}$ has period $1/T$, it is sufficient to choose $\theta\sim U(0,1/T)$.

\section{Maximum Likelihood Phase Estimation}

As is introduced, UPEA shows more of its power when we allow repeating it for several times.
Suppose we have repeated PEA or UPEA for $R$ times, and get a set of estimation $\{\tilde{\varphi}_1,\tilde{\varphi}_2,\cdots,\tilde{\varphi}_R\}$.
The estimation $\tilde{\varphi}$ is obtained by maximizing the likelihood function,
\begin{equation}
    L(\tilde{\varphi};\tilde{\varphi}_1,\cdots,\tilde{\varphi}_R) = \prod_{j=1}^R \left(\frac{\sin(T\pi(\tilde{\varphi}_j-\tilde{\varphi}))}{T\sin(\pi(\tilde{\varphi}_j-\tilde{\varphi}))}\right)^2.
    \label{eq:likelihood}
\end{equation}

\begin{figure*}
    \centering
    \subfloat[The bias calculated by simulating for $2^{16}$ times.]{
        \includegraphics[width=.45\textwidth]{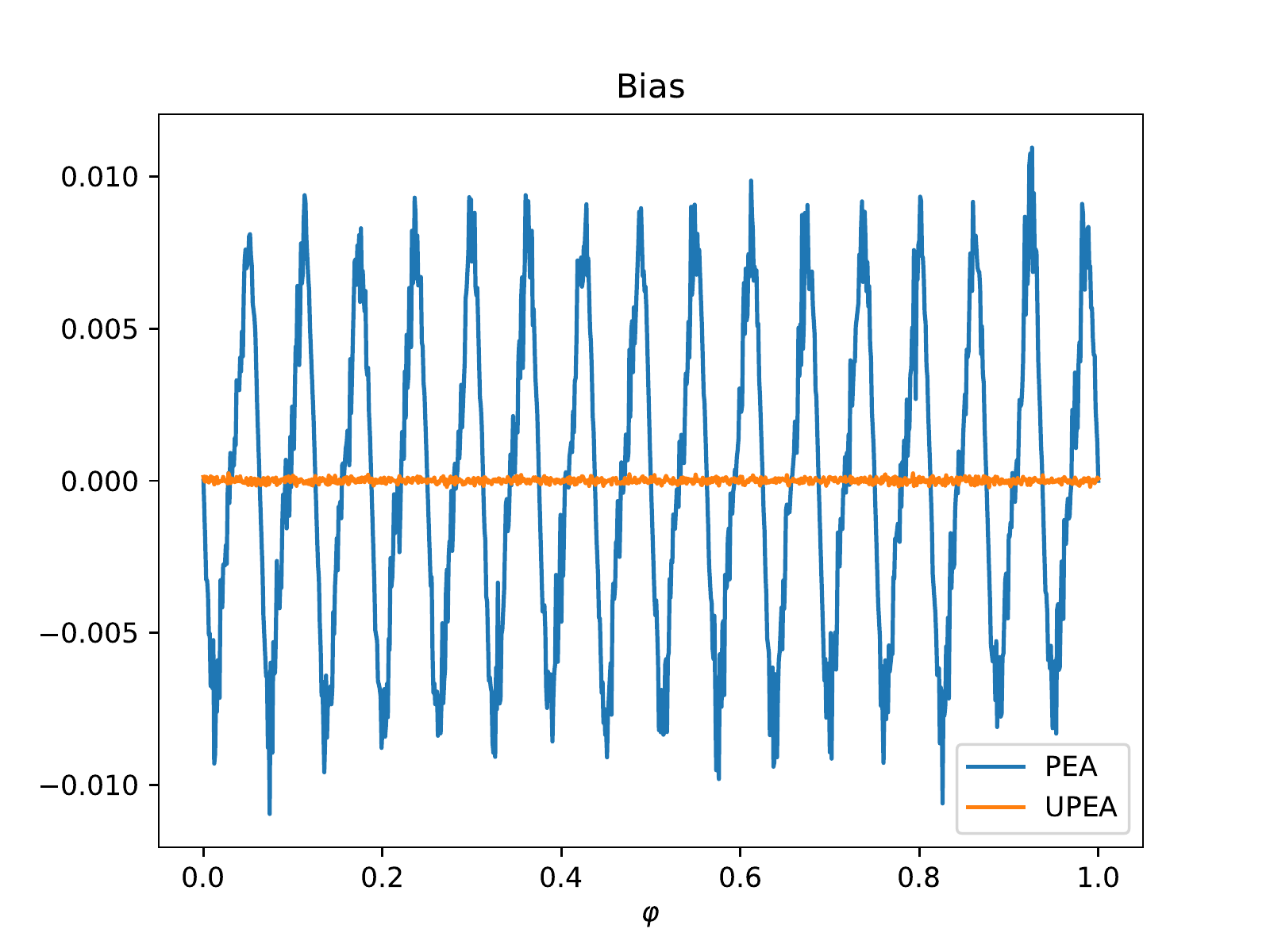}
        \label{fig:bias-pp}
    }\quad
    \subfloat[The MAE calculated by simulating for $2^{16}$ times.]{
        \includegraphics[width=.45\textwidth]{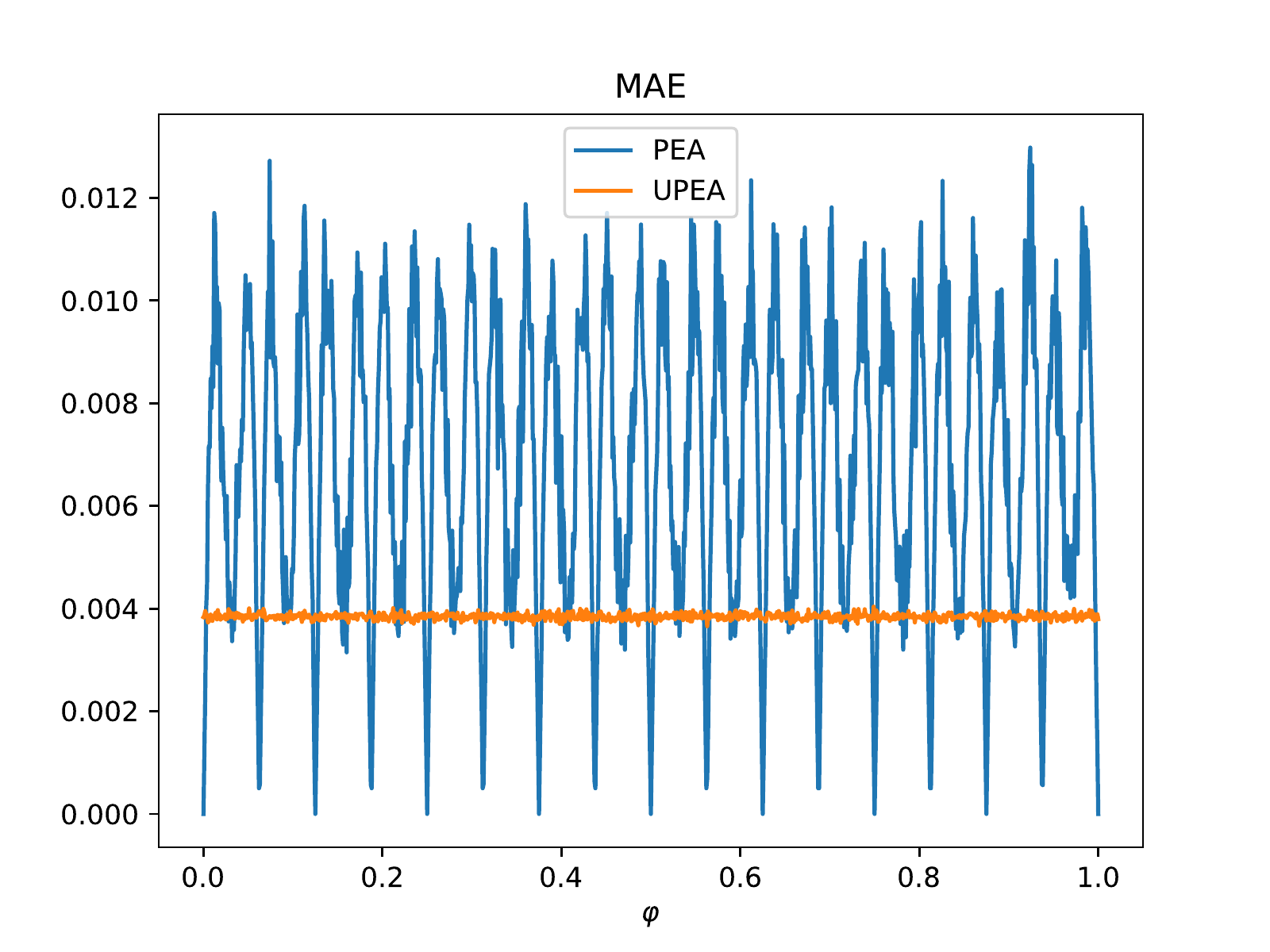}
        \label{fig:mae-pp}
    }
    \caption{A comparison between PEA and UPEA, using maximum likelihood estimation, with parameters $T=R=16$.}
    \label{fig:bias-mae-pp}
\end{figure*}

\begin{figure}
    \centering
    \includegraphics[width=.45\textwidth]{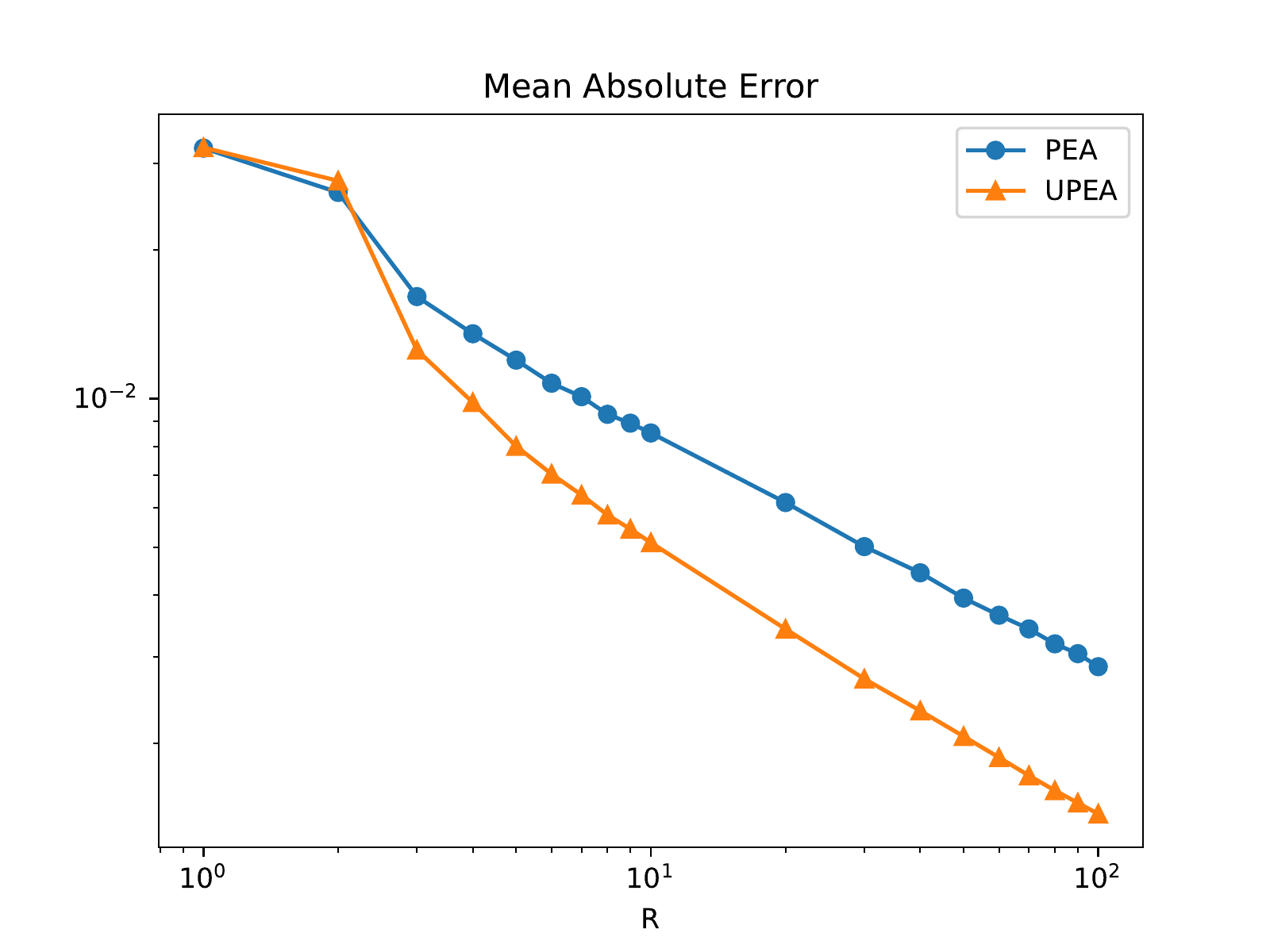}
    \caption{The error behavior with respect to $R$, where $T=16$. }
    \label{fig:mae-rep}
\end{figure}

We study the bias and MAE of PEA and UPEA with $R=T=16$, and the results are shown in \autoref{fig:bias-mae-pp}\autoref{sub@fig:bias-pp}\autoref{sub@fig:mae-pp}.
We also fix $T=16$ and study the bias and MAE behavior with respect to $R$, as shown in \autoref{fig:mae-rep}.
In summary,
\begin{itemize}
    \item The maximum likelihood estimation algorithm maintains the unbiasedness of UPEA;
    \item For $R\geq 3$, there is a sudden decrement of error for both PEA and UPEA;
    \item For $R\geq 3$, UPEA has a smaller MAE than PEA.
\end{itemize}

We conclude that UPEA behaves even better than PEA when combined with the maximum likelihood estimation.

\section{Application in Quantum Counting}

The quantum counting algorithm (QCA) is an important application of PEA.
Given a Boolean function $f:\{0,1,\cdots,N-1\}\rightarrow\{0,1\}$, where $N=2^n(n\in\mathbb{Z}_+)$, the quantum counting algorithm can estimate the quantity,
\begin{equation}
    M = \sum_{j=0}^{N-1} f(j).
\end{equation}

The key idea of quantum counting is that the uniform superposition state,
\begin{equation}
    \ket{u} = \frac{1}{\sqrt{N}}\sum_{j=0}^{N-1}\ket{j},
\end{equation}
is on a plane spanned by two eigenvectors $\ket{\psi_\pm}$ of the Grover's iteration~\cite{Grover1997,qcqi} $G_f$ for Boolean function $f$, with eigenvalues $e^{\pm 2\pi i\varphi}$, where
\begin{equation}
    M = N\sin^2(\pi\varphi).
\end{equation}

To be specific,
\begin{equation}
    \ket{u} = \frac{1}{\sqrt{2}}\left(e^{i\pi\varphi}\ket{\psi_+}+e^{-i\pi\varphi}\ket{\psi_-}\right).
\end{equation}

Applying PEA to $G_f$ and $\ket{\psi}$, the output $\tilde{\varphi}=s/T$ obeys the distribution $P(\tilde{\varphi}|\varphi)$ or $P(\tilde{\varphi}|-\varphi)$ with equal probability.
Finally, the result of QCA is,
\begin{equation}
    \tilde{M} = N\sin^2(\pi\tilde{\varphi}).
\end{equation}

If we replace the PEA step in quantum counting with UPEA, then $s/T$ is an estimation of $\varphi+\theta$ or $-\varphi+\theta$ with equal probability.
Similarly, the estimation is given by
\begin{equation}
    \tilde{M} = N\sin^2\left[\pi\left(\frac{s}{T}-\theta\right)\right].
    \label{eq:UQCA-old}
\end{equation}

\begin{figure*}
    \centering
    \subfloat[The bias calculated by simulating for $2^{16}$ times.]{
        \includegraphics[width=.45\textwidth]{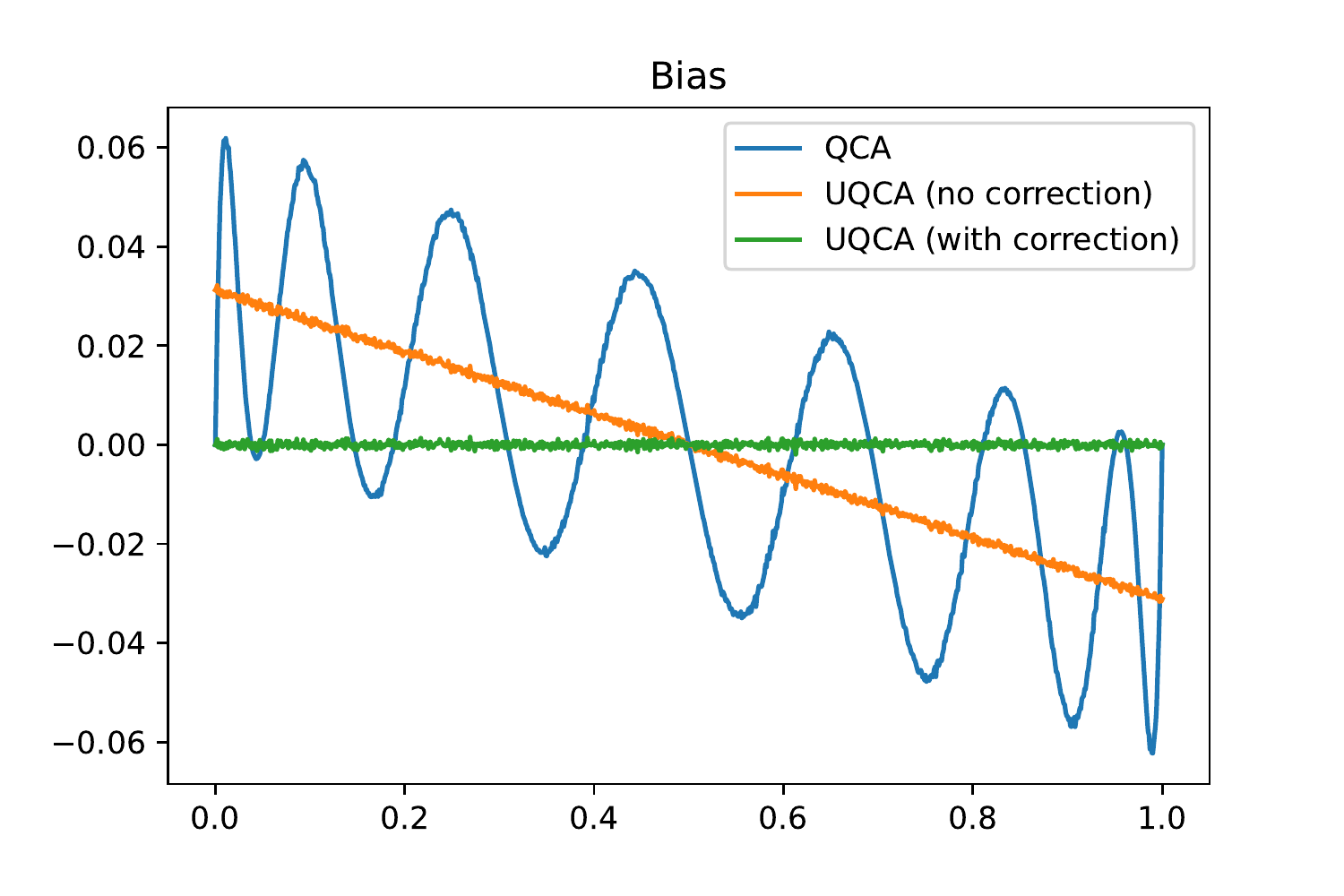}
    }\quad
    \subfloat[The MAE calculated by simulating for $2^{16}$ times.]{
        \includegraphics[width=.45\textwidth]{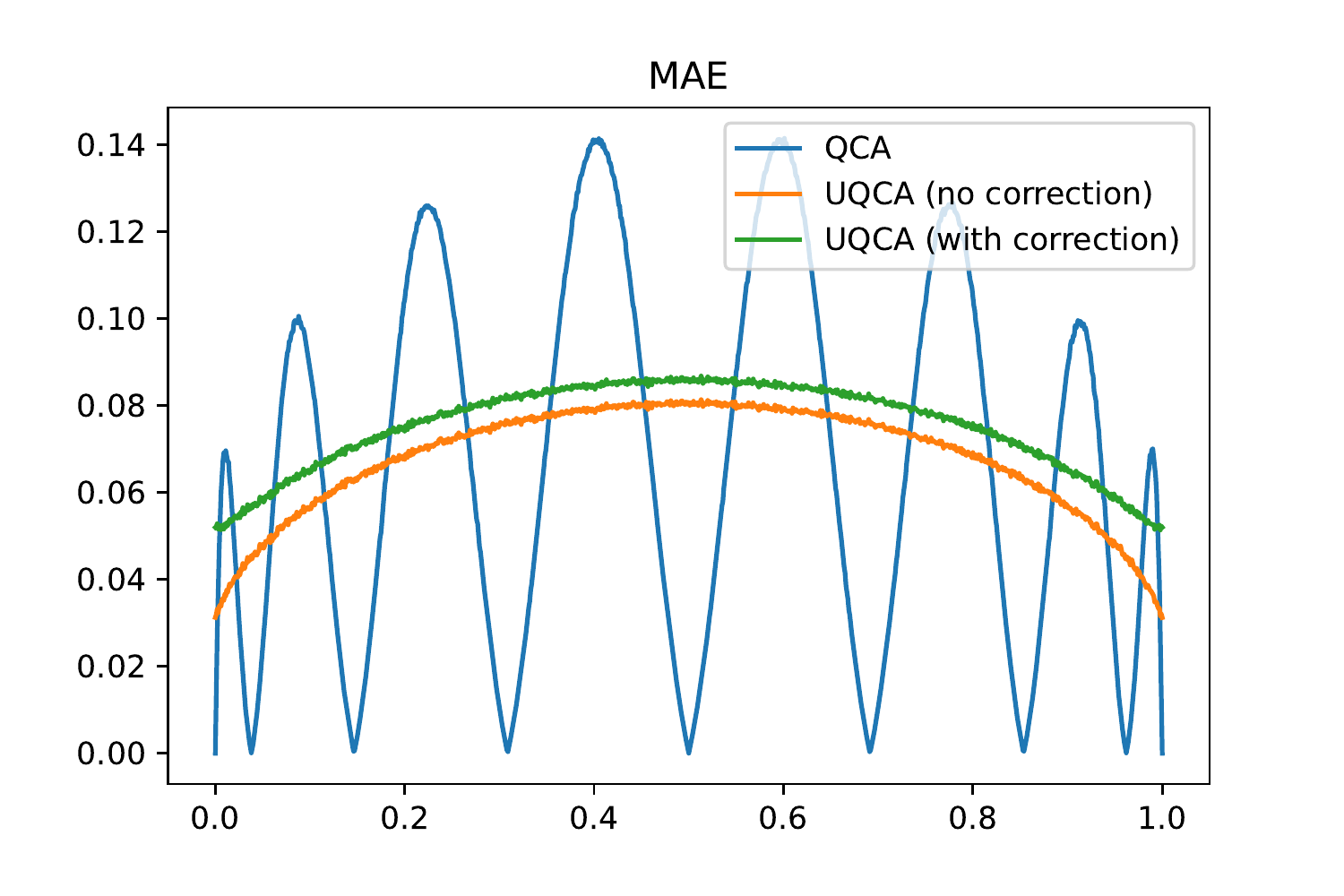}
    }
    \caption{The bias and MAE of quantum counting, with parameter $T=16$. The $x$-axis stands for $m$, and the $y$-axis stands for bias or MAE.}
    \label{fig:bias_mae_uqca}
\end{figure*}

For convenience, we define $m=M/N$ and $\tilde{m}=\tilde{M}/N$.
Though our estimation about $\varphi$ or $-\varphi$ is unbiased, the nonlinear mapping from $\varphi$ to $m$ will break the unbiasedness, as confirmed by our simulation experiments in \autoref{fig:bias_mae_uqca}.
Here the bias and MAE is defined as,
\begin{align}
    B_{QCA}(m) &= \sum_{\tilde{M}} (\tilde{m}-m) P_{QCA}(\tilde{m}|m),\\
    M_{QCA}(m) &= \sum_{\tilde{M}} |\tilde{m}-m| P_{QCA}(\tilde{m}|m),
\end{align}
where $P_{QCA}(\tilde{m}|m)=P_{PEA}(\tilde{\varphi}|\varphi)$.
In \autoref{fig:bias_mae_uqca}, the quantity $B_{UQCA}(m)$ shows a linear relationship with $m$.

Theoretically, the bias of UQCA is,
\begin{align}
    & B_{UQCA}(m) \nonumber
    \\ = & \int_0^1 (\tilde{m}-m)\rho_{UQCA}(\tilde{m}|m) \operatorname{d}\tilde{m}
    \\ = & \int_0^1 [\sin^2(\pi\tilde{\varphi})-\sin^2(\pi\varphi)] \rho_{UPEA}(\tilde{\varphi}|\varphi)\operatorname{d}\tilde{\varphi},
\end{align}
where $\rho_{UQCA}(\tilde{m}|m)$ is the probability density function of obtaining an estimation $\tilde{m}$ using UQCA when the ground truth is $m$, and $\rho_{UPEA}(\tilde{\varphi}|\varphi)$ is the probability density function of obtaining an estimation $\tilde{\varphi}$ using UPEA when the ground truth is $\varphi$.

Using the period-1 property,
\begin{align}
    & B_{UQCA}(m) \nonumber
    \\ = & \int_0^1 [\sin^2(\pi\tilde{\varphi})-\sin^2(\pi\varphi)] \frac{\sin^2(T\pi(\tilde{\varphi}-\varphi))}{T\sin^2(\pi(\tilde{\varphi}-\varphi))}\operatorname{d}\tilde{\varphi}
    \\ = & \frac{1}{T} \int_{-1/2}^{1/2} [\sin^2(\pi(\varphi+\phi))-\sin^2(\pi\varphi)] \frac{\sin^2(T\pi\phi)}{T\sin^2(\pi\phi)}\operatorname{d}\phi
    \\ = & \frac{1}{T} \cos(2\pi\varphi) \int_0^1 \sin^2(T\pi\phi)\operatorname{d}\phi
    \\ = & \frac{1}{2T} \cos(2\pi\varphi)
    \\ = & \frac{1-2m}{2T}.
\end{align}

Therefore, QCA cannot be made unbiased by simply replacing PEA with UPEA.
Indeed, we can output
\begin{equation}
    m' = \left(1-\frac{1}{T}\right)^{-1}\left(\tilde{m}-\frac{1}{2T}\right),
    \label{eq:correction-uqca}
\end{equation}
instead of $\tilde{m}$ to make it unbiased.
But such correction step can bring a little cost in MAE, as shown in \autoref{fig:bias_mae_uqca}, since $m'$ is amplified $(1-1/2T)^{-1}$ times in Eq.~\eqref{eq:correction-uqca}.

Similarly, if we repeat UPEA for $R$ times and obtain $\{\tilde{\varphi}_1,\tilde{\varphi}_2,\cdots,\tilde{\varphi}_R\}$, we can use maximum likelihood estimation to make the result more robust.

In UQCA, the bias is,
\begin{align}
    & B_{UQCA}(m) \nonumber
    \\ = & \int_0^1 \rho_{UPEA}(\tilde{\varphi}_1;\varphi)\operatorname{d}\tilde{\varphi}_1 \cdots \int_0^1 \rho_{UPEA}(\tilde{\varphi}_R;\varphi)\operatorname{d}\tilde{\varphi}_R \nonumber
    \\ & \cdot \left[\sin^2\left(\pi\argmax_{\varphi'}L(\varphi';\tilde{\varphi}_1,\cdots,\tilde{\varphi}_R)\right) - \sin^2(\pi\varphi)\right]
    \\ = & \int_0^1 \rho_{UPEA}(\tilde{\varphi}_1-\varphi;0)\operatorname{d}\tilde{\varphi}_1 \cdots \int_0^1 \rho_{UPEA}(\tilde{\varphi}_R-\varphi;0)\operatorname{d}\tilde{\varphi}_R  \nonumber
    \\ & \cdot \left[\sin^2\left(\pi\argmax_{\varphi'}L(\varphi';\tilde{\varphi}_1,\cdots,\tilde{\varphi}_R)\right) - \sin^2(\pi\varphi)\right]
    \\ = & \int_0^1 \rho_{UPEA}(\tilde{\varphi}_1;0)\operatorname{d}\tilde{\varphi}_1 \cdots \int_0^1 \rho_{UPEA}(\tilde{\varphi}_R;0)\operatorname{d}\tilde{\varphi}_R  \nonumber
    \\ & \cdot \left[\sin^2\left(\pi\argmax_{\varphi'}L(\varphi';\tilde{\varphi}_1+\varphi,\cdots,\tilde{\varphi}_R+\varphi)\right) - \sin^2(\pi\varphi)\right]
    \\ = & \int_0^1 \rho_{UPEA}(\tilde{\varphi}_1;0)\operatorname{d}\tilde{\varphi}_1 \cdots \int_0^1 \rho_{UPEA}(\tilde{\varphi}_R;0)\operatorname{d}\tilde{\varphi}_R  \nonumber
    \\ & \cdot \left[\sin^2\left(\pi\varphi+\pi\argmax_{\varphi'}L(\varphi';\tilde{\varphi}_1,\cdots,\tilde{\varphi}_R)\right) - \sin^2(\pi\varphi)\right]
    \\ = & \cos(2\pi\varphi) \int_0^1 \rho_{UPEA}(\tilde{\varphi}_1;0)\operatorname{d}\tilde{\varphi}_1 \cdots \int_0^1 \rho_{UPEA}(\tilde{\varphi}_R;0)\operatorname{d}\tilde{\varphi}_R  \nonumber
    \\ & \cdot \sin^2\left(\pi\argmax_{\varphi'}L(\varphi';\tilde{\varphi}_1,\cdots,\tilde{\varphi}_R)\right)
    \\ = & B_{UQCA}(0) \cos(2\pi\varphi)
    \\ = & B_{UQCA}(0) (1-2m)
\end{align}

\begin{figure*}
    \centering
    \subfloat[The bias calculated by simulating for $2^{16}$ times.]{
        \includegraphics[width=.45\textwidth]{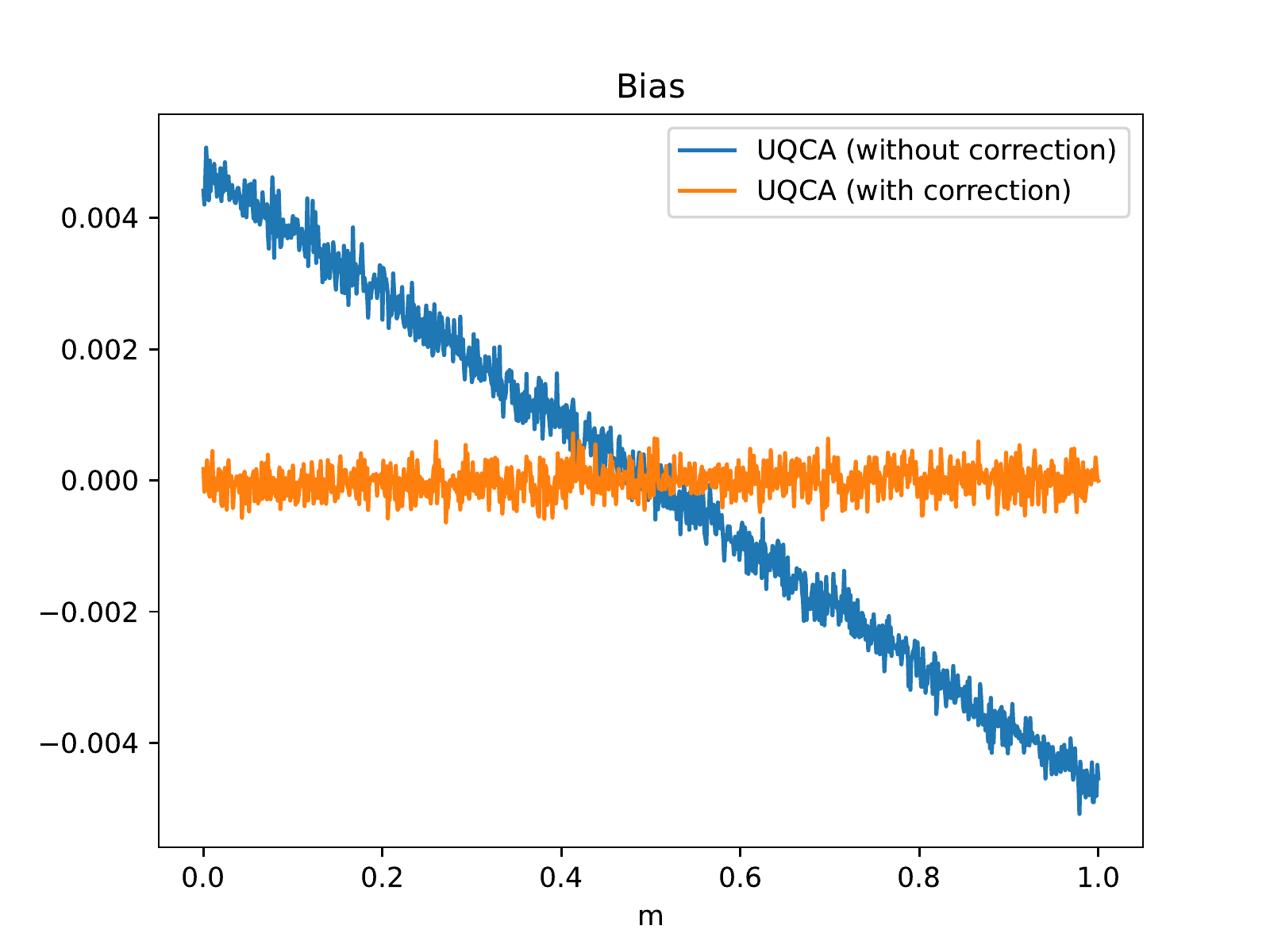}
    }\quad
    \subfloat[The MAE calculated by simulating for $2^{16}$ times.]{
        \includegraphics[width=.45\textwidth]{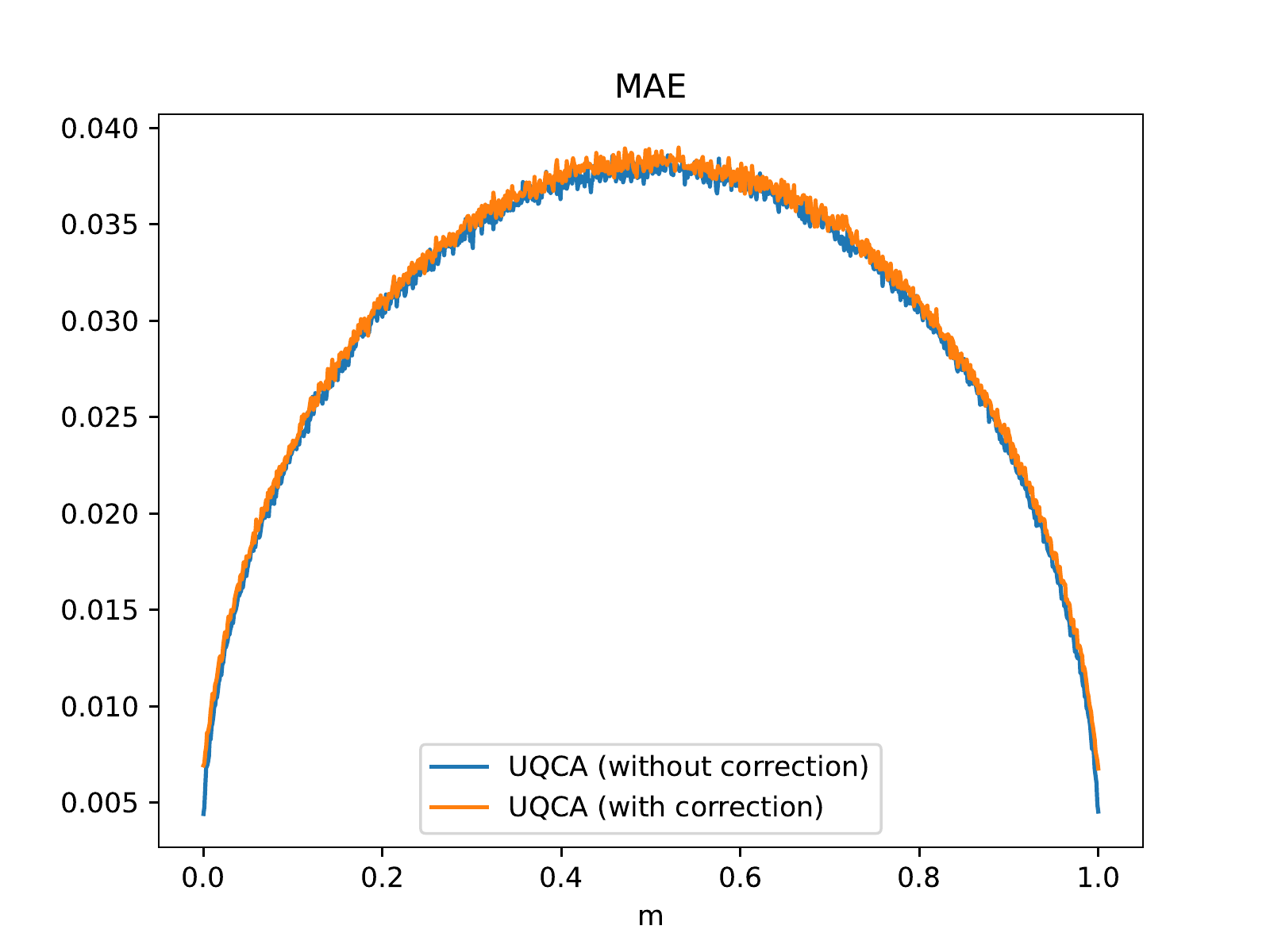}
    }
    \caption{The bias and MAE of quantum counting, with parameter $T=16$ and $R=3$.}
    \label{fig:correction-muqca}
\end{figure*}

Thus, the bias of UQCA with $R$ repetitions is also linear about $m$.
The only thing left for the correction formula is $B_{UQCA}(0)$, which can be pre-calculated by simulation.
Before performing correction for UQCA with parameters $T$ and $R$, one should first run simulations to calculate the corresponding $b:=B_{UQCA}(0)$, then the correction formula is,
\begin{equation}
    m' = \frac{\tilde{m}-b}{1-2b},
    \label{eq:correction-muqca}
\end{equation}
where $\tilde{m}$ is the result of maximum likelihood estimation, and $m'$ is the correction output.

For example, by simulating UQCA with $m=0,T=16,R=3$ for $2^{16}$ times we get $b\approx0.004775$, then Eq.~\eqref{eq:correction-muqca} can be applied to make UQCA unbiased.
We do experiments to compare UQCA with or without correction, as shown in \autoref{fig:correction-muqca}.

\begin{figure}
    \centering
    \includegraphics[width=.45\textwidth]{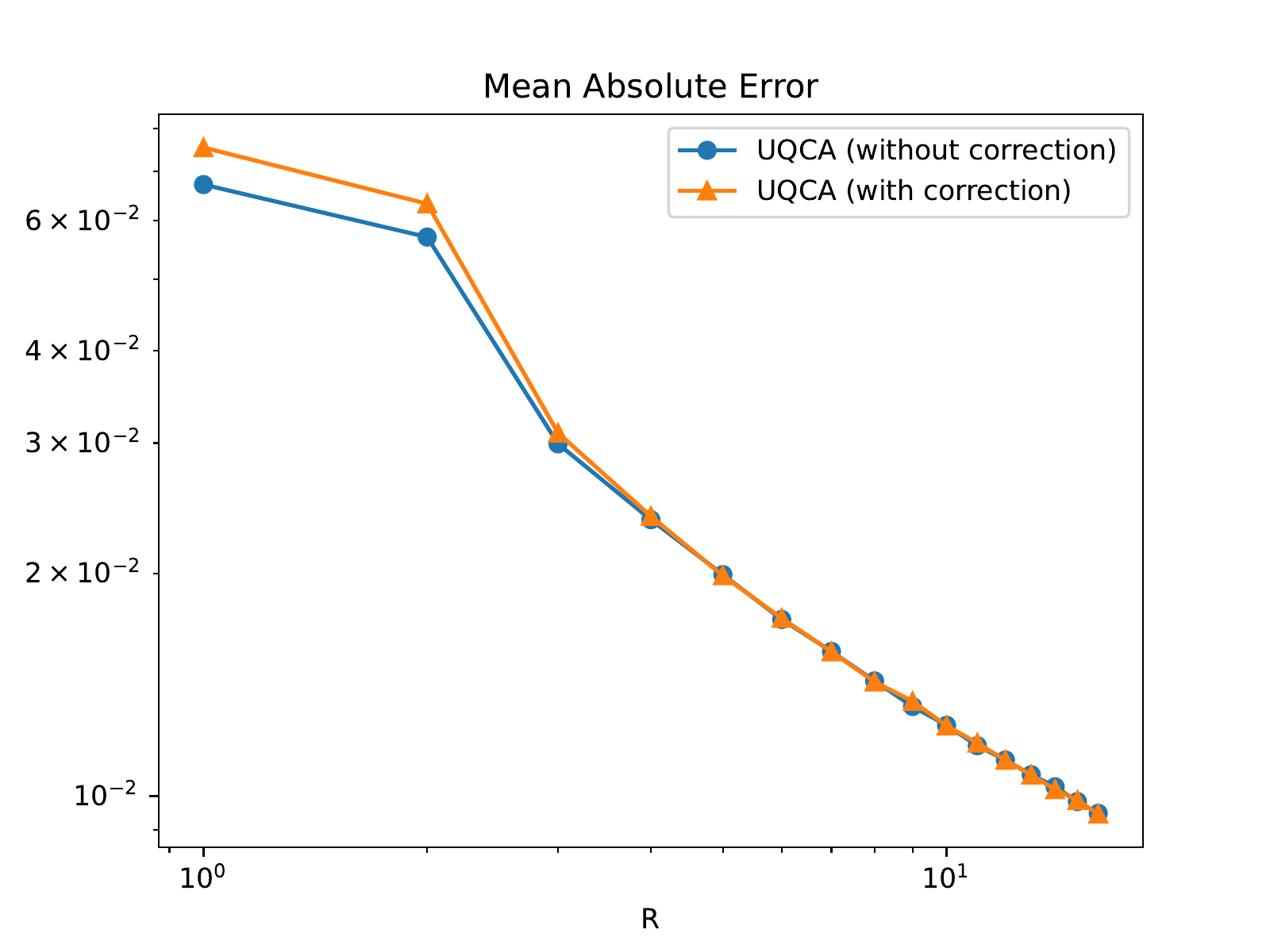}
    \caption{The MAE of UQCA with or without correction for $T=16$ and different $R$.}
    \label{fig:fig:correction-muqca-R}
\end{figure}

We also do experiments for UQCA with or without correction for $T=16$ and different $R$.
The results in \autoref{fig:fig:correction-muqca-R} shows that the extra error brought by the correction decays as $R$ grows.

\section{Conclusion}

The original form of phase estimation algorithm suffers a periodical bias, which prevents its accuracy from reaching an arbitrarily low level.
We propose an unbiased phase estimation algorithm, by introducing a uniformly distributed variable $\theta$ to the original phase estimation algorithm.
We also show that a maximum likelihood estimation post-processing step can make UPEA more robust when $R\geq 3$, as it keeps the unbiasedness and reduce the MAE quickly.

Finally, we apply UPEA to quantum counting.
We point out that a direct substitution of UPEA for PEA cannot make quantum counting unbiased, and there is a linear relationship between the bias and $m$, the ground truth of quantum counting.
By applying a correction step, the bias can vanish, with an extra cost of MAE in the meantime.
Moreover, by repeating for $R$ times and using maximum likelihood estimation, we prove that the linear relationship still holds, and the extra cost decays quickly while the unbiasedness is maintained.

\bibliographystyle{plain}
\bibliography{ref}

\end{document}